\documentclass[pdftex,
twocolumn,
epjc3]
{svjour3_own}
\RequirePackage[T1]{fontenc}
\smartqed  % flush right qed marks, e.g. at end of proof
\RequirePackage{flushend}
\RequirePackage[numbers,sort&compress]{natbib}
\RequirePackage[colorlinks,citecolor=blue,urlcolor=blue,linkcolor=blue]{hyperref}
\usepackage{graphicx}% Include figure files
\usepackage{dcolumn}% Align table columns on decimal point
\usepackage{bm}% bold math
\usepackage[mathlines]{lineno}% Enable numbering of text and display math
\usepackage{siunitx}
\usepackage{color}
\usepackage{footnote}
\usepackage[export]{adjustbox}
\makesavenoteenv{tabular}
\makesavenoteenv{table}

\def\bsq#1{%both single quotes
	\lq{#1}\rq}

\journalname{Eur. Phys. J. C}

\begin{document}
	\title{Search for neutrinos from decaying dark matter with IceCube}
	
	\onecolumn
	\author{IceCube Collaboration: M.~G.~Aartsen\thanksref{Christchurch}
		\and M.~Ackermann\thanksref{Zeuthen}
		\and J.~Adams\thanksref{Christchurch}
		\and J.~A.~Aguilar\thanksref{BrusselsLibre}
		\and M.~Ahlers\thanksref{Copenhagen}
		\and M.~Ahrens\thanksref{StockholmOKC}
		\and I.~Al~Samarai\thanksref{Geneva}
		\and D.~Altmann\thanksref{Erlangen}
		\and K.~Andeen\thanksref{Marquette}
		\and T.~Anderson\thanksref{PennPhys}
% ...
	}
	\authorrunning{IceCube Collaboration}
	\thankstext{a}{Earthquake Research Institute, University of Tokyo, Bunkyo, Tokyo 113-0032, Japan}
	\institute{
% ...
Universit\'e Libre de Bruxelles, Science Faculty CP230, B-1050 Brussels, Belgium \label{BrusselsLibre}
% ...
\and Dept.~of Physics and Astronomy, University of Canterbury, Private Bag 4800, Christchurch, New Zealand \label{Christchurch}
% ...
\and Niels Bohr Institute, University of Copenhagen, DK-2100 Copenhagen, Denmark \label{Copenhagen}
% ...
\and Erlangen Centre for Astroparticle Physics, Friedrich-Alexander-Universit\"at Erlangen-N\"urnberg, D-91058 Erlangen, Germany \label{Erlangen}
		\and D\'epartement de physique nucl\'eaire et corpusculaire, Universit\'e de Gen\`eve, CH-1211 Gen\`eve, Switzerland \label{Geneva}
% ...
\and Department of Physics, Marquette University, Milwaukee, WI, 53201, USA \label{Marquette}
% ...
\and Oskar Klein Centre and Dept.~of Physics, Stockholm University, SE-10691 Stockholm, Sweden \label{StockholmOKC}
% ...
\and Dept.~of Physics, Pennsylvania State University, University Park, PA 16802, USA \label{PennPhys}
% ...
\and DESY, D-15738 Zeuthen, Germany \label{Zeuthen}
	}

	\maketitle
	\date{Received: 10-Apr-2018 / Accepted: 21-Sep-2018}
	\twocolumn
	
	\begin{abstract}
		With the observation of high-energy astrophysical neutrinos by the IceCube Neutrino Observatory, interest has risen in models of PeV-mass decaying dark matter particles to explain the observed flux. We present two dedicated experimental analyses to test this hypothesis. 
		One analysis uses six years of IceCube data focusing on muon neutrino \bsq{track} events from the Northern Hemisphere, while the second analysis uses two years of \bsq{cascade} events from the full sky. Known background components and the hypothetical flux from unstable dark matter are fitted to the experimental data. Since no significant excess is observed in either analysis, lower limits on the lifetime of dark matter particles are derived: We obtain the strongest constraint to date, excluding lifetimes shorter than $10^{28}\si{\s}$ at 90\% CL for dark matter masses above $\SI{10}{\tera\electronvolt}$.
	\end{abstract}
	
	
	\section{\label{sec:introduction} High-Energy Neutrinos and Dark Matter Decay}
	To this day, the origin of the flux of high-energy neutrinos discovered by IceCube~\cite{PhysRevLett.113.101101, Aartsen:2016xlq} remains unidentified~\cite{PS7Year:2017}. Likewise, the nature and properties of dark matter (DM) are among the most important open questions in physics. If the hypothetical dark matter particles are unstable on time-scales longer than the age of the universe, then the two questions may be linked~\cite{Ellis:1990nb,Gondolo:1991rn}, i.e. neutrinos produced in dark matter decays could contribute to the observed astrophysical flux. Following the IceCube discovery of cosmic neutrinos up to $\si{\peta\eV}$ energies, there has been renewed interest in this possibility~\cite{1475-7516-2010-04-017,Esmaili:2014rma, Rott:2014kfa, Bai:2013nga, Esmaili:2013gha, Bhattacharya:2014vwa, Esmaili:2012us, Fong:2014bsa, Aisati:2015vma, Boucenna:2015tra, Troitsky:2015cnk, Chianese:2016opp, Chianese:2016kpu, Bhattacharya:2017jaw, Chianese:2017nwe}. In particular, the connection between neutrinos and gamma-rays from DM decay has been discussed in further detail~\cite{Murase:2012xs, Cohen:2016uyg,Esmaili:2015xpa, Murase:2015gea, Blanco:2017sbc, Kalashev:2016cre, Kuznetsov:2016fjt, Kachelriess:2018rty, Ema:2013nda, Ema:2014ufa, Anchordoqui:2015lqa, Zavala:2014dla}.
	%This has been studied in phenomenological analysis in Refs.~\cite{1475-7516-2010-04-017,Esmaili:2014rma, Rott:2014kfa, Bai:2013nga, Esmaili:2013gha, Bhattacharya:2014vwa, Esmaili:2012us, Fong:2014bsa, Aisati:2015vma, Boucenna:2015tra, Troitsky:2015cnk, Chianese:2016opp, Chianese:2016kpu, Bhattacharya:2017jaw, Chianese:2017nwe}, explicit models have been proposed in Refs.~\cite{Higaki:2014dwa, Daikoku:2015vsa, Roland:2015yoa, Ko:2015nma, Dev:2016uxj, Fiorentin:2016avj, DiBari:2016guw, Chianese:2016smc, Borah:2017xgm, Hiroshima:2017hmy, Sahoo:2017cqg, Chakravarty:2017hcy, Dhuria:2017ihq, Dev:2016qbd}. Finally, the connection between neutrino and gamma-ray observations from DM decay has been discussed in Refs~\cite{Murase:2012xs, Cohen:2016uyg,Esmaili:2015xpa, Murase:2015gea, Blanco:2017sbc, Kalashev:2016cre, Kuznetsov:2016fjt, Kachelriess:2018rty, Ema:2013nda, Ema:2014ufa, Anchordoqui:2015lqa, Zavala:2014dla}.
	
	We present two dedicated analyses to test whether the description of the observed neutrino flux can be improved by an additional component from heavy ($m_{\textrm{DM}} > \SI{10}{\TeV}$) dark matter decays as an alternative to bottom-up scenarios of astrophysical acceleration~\cite{doi:10.1146/annurev-nucl-101916-123304}. 
	Such heavy particles are receiving increased attention because the classic WIMP paradigm of weak-scale mass dark matter is disfavoured by the negative results in searches for new physics at the LHC~\cite{Buchmueller2017}, in direct DM detection experiments~\cite{0954-3899-43-1-013001,Aprile:2017iyp,Akerib:2016vxi,Cui:2017nnn,Amole:2017dex}, and in searches for DM annihilation into neutrinos~\cite{Aartsen:2017ulx, Albert:2016emp} or gamma-rays~\cite{Ahnen:2016qkx,PhysRevLett.117.111301,HAWC_dm,HAWC_dm_GC_2018,HAWC:2018eaa}.
	
	Our results significantly improve upon the best previous experimental bounds on decaying dark matter obtained with gamma rays~\cite{Ackermann:2012rg,HAWC_dm,HAWC_dm_GC_2018,HAWC:2018eaa}, neutrinos~\cite{Abbasi:2011eq}, and those derived from high-energy cosmic rays and the cosmic microwave background radiation~\cite{Ellis:1990nb,Gondolo:1991rn}.
	
	\section{IceCube Detector and Event Selections}
	
	IceCube is a cubic-kilometer ice Cherenkov detector located at the South Pole, situated between 1450\,m and 2450\,m below the surface~\cite{1748-0221-12-03-P03012}. Charged particles produced in neutrino interactions with the Antarctic ice or the bedrock below are detected by the Cherenkov light they emit, allowing the reconstruction of the originating neutrino's direction and energy~\cite{1748-0221-9-03-P03009}.
	
	The presented analyses use two different event samples. The first analysis is based on six years of $\nu_{\mu}$ charged-current data collected between 2009 and 2015, i.e., track-like events from the Northern Hemisphere. More details can be found in Ref.~\cite{Aartsen:2016xlq}. 
	The second analysis uses two years of data collected between 2010 and 2012. The event selection is based on a previous study~\cite{Aartsen:2014muf}, modified to select only cascade events from the full sky which are produced in NC interactions or CC interactions of $\nu_{e}$ or $\nu_{\tau}$. 
	Note that in the following no distinction is made between particles and anti-particles; the labels \textit{neutrino} and \textit{lepton} include the respective anti-particles and the used cross-sections incorporate both particles and anti-particles.
	
	The two analysis samples are statistically independent, and while the track sample contains a much larger number of events, the full-sky coverage and better energy resolution of the cascade sample (see Table \ref{tab:event_selections}) lead to comparable sensitivities.
	
	\begin{table}[ht]%[bhtp]
		\centering
		\caption{\label{tab:event_selections} Summary of the two event samples. Detailed sample descriptions can be found in Refs.~\cite{Aartsen:2016xlq, Aartsen:2014muf}.}
		\begin{tabular*}{\columnwidth}{@{\extracolsep{\fill}}lcc}
			\hline
			&\textrm{Tracks}& \textrm{Cascades}\\
			\hline
			\textrm{Number of events} & 352,294 & 278\\
			\textrm{Livetime} & 2060 days & 641 days\\ %(2009-2015)  (2010-2012)\\
			\textrm{Sky coverage} & \textrm{North} ($\textrm{zenith}>85^\circ$) & \textrm{Full Sky}\\
			\textrm{Atm. muon background} & 0.3\% & 10\% \\
			\textrm{Median reconstr. error} & $<0.5^{\circ} (\textrm{E}_{\nu}>\SI{100}{\tera\eV})$ & $ \sim 10^{\circ}$ \\
			\textrm{Energy uncertainty} & $ \sim 100\%$ & $\sim 10\%$ \\
			\hline
		\end{tabular*}
	\end{table}
	
	\section{Analysis}
	To test whether the observed flux of high-energy neutrinos (partly) arises from heavy decaying dark matter, a forward-folding likelihood fit of the distribution of reconstructed energy and direction is performed on both datasets, similar to Refs.~\cite{Aartsen:2016xlq} and~\cite{Aartsen:2014muf}.
	The total observed flux is modelled as a sum of background and signal flux components. Each of these components is described by a parametrized flux template that depends on the fitted model parameters.
	
	\subsection{Flux components}
	
	Cosmic-ray air showers produce secondary mesons which decay into charged leptons and neutrinos. These atmospheric neutrinos are the main source of background in both data samples. They can  be further divided into \textit{conventional} atmospheric neutrinos produced by the decay of pions and kaons and \textit{prompt} neutrinos produced by the decay of charmed mesons. This latter flux is sub-dominant at high energies and has not been separately identified yet~\cite{Aartsen:2016xlq}. Atmospheric neutrino flux predictions are taken from Refs.~\cite{PhysRevD.75.043006} and~\cite{PhysRevD.78.043005} for the conventional (modified to account for the cosmic-ray knee~\cite{Aartsen:2016xlq}) and prompt component, respectively. 
	From the Southern Hemisphere, cosmic-ray induced atmospheric muons can also penetrate the ice, reach the detector and mimic a neutrino signal. After application of appropriate event selections, the atmospheric muon contamination is negligible in the track-like sample and below 10\% in the cascade sample.
	
	Astrophysical neutrinos from cosmic rays interacting in or near their production sites constitute a second background flux to the targeted signal of neutrinos from decaying dark matter.
	Since the origin of cosmic rays is unknown, an exact modelling of this astrophysical flux is not possible. A generic parametrization of these astrophysical neutrinos as an isotropic flux with a power-law energy spectrum agrees well with present measurements~\cite{PhysRevLett.113.101101,Aartsen:2016xlq} and is therefore used in the fitting. The spectral index $\gamma$ and the flux normalization $\Phi_\textrm{astro}$ are taken as free parameters in the fit.
	
\begin{figure}[ht]
		\includegraphics[width=.48\textwidth,left]{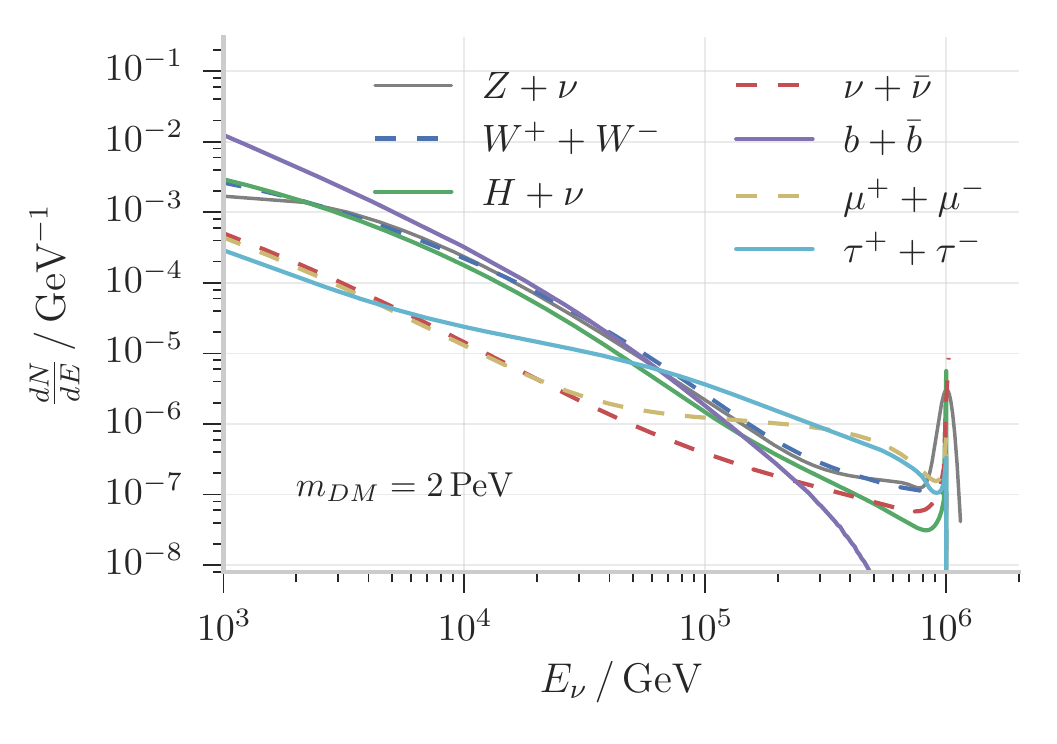}  
		\protect\caption{Neutrino yield per decay as a function of neutrino energy (flavour-averaged): All considered decay channels ($\textrm{BR}=100\%$) are presented for an assumed dark matter mass of $\SI{2}{\peta\electronvolt}$.}
		\label{fig:decay_spectra}
	\end{figure}

	When heavy dark matter decays into standard model particles, neutrinos are necessarily expected in the final state~\cite{Sarkar:2001se}. Observing these neutrinos would thus constitute an indirect probe of the scenario of decaying dark matter. The energy spectrum, $dN_{\nu}/dE_{\nu}$, of the expected neutrinos depends on the exact decay mechanism and is model dependent. In this analysis, several ``hard'' (e.g., dark matter decaying directly into neutrinos~\cite{Feldstein:2013kka,Rott:2014kfa,Covi:2008jy}) and ``soft'' (e.g., neutrinos produced in the subsequent hadronic decay-chain of standard model particles~\cite{1475-7516-2010-04-017}) decays are used as benchmark channels. Their spectra were simulated with PYTHIA 8.1~\cite{Sjostrand:2007gs} and are shown in Fig.~\ref{fig:decay_spectra}.
	
\begin{figure}[ht]%[ht]
		\includegraphics[width=.48\textwidth,left]{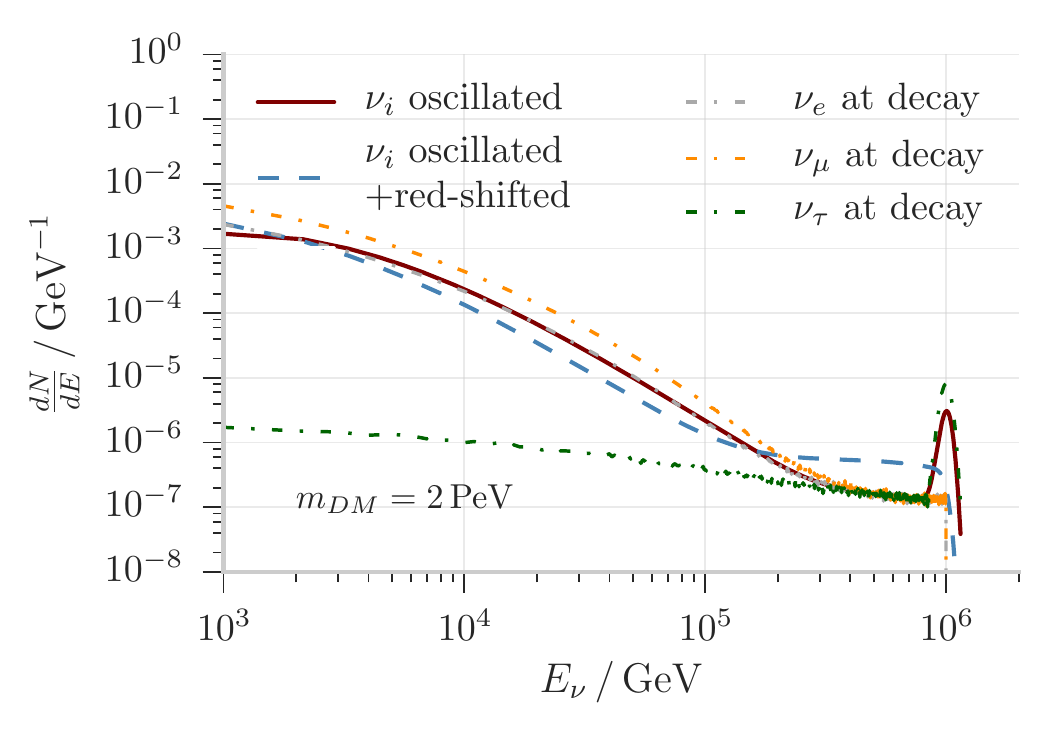}  
		\protect\caption{Neutrino yield per decay as a function of neutrino energy assuming the hard decay channel $DM\rightarrow Z+\nu_{\tau}$: the effects of neutrino mixing and red-shift are illustrated.}
		\label{fig:decay_spectra_details}
	\end{figure}

	At Earth, the neutrino flux from dark matter decays has to be subdivided into a galactic and an extragalactic component. The expected energy distribution of the galactic component $\Phi^{\textrm{Halo}}$ follows the initial decay spectrum. Its angular distribution incorporates the (uncertain) distribution of dark matter in the Milky Way halo via the line-of-sight integral~\cite{PalomaresRuiz:2007ry}. The Burkert halo profile~\cite{Burkert:1995yz} with best-fit parameters from Ref.~\cite{Nesti:2013uwa} is used as a benchmark and other halo profiles are considered as systematic uncertainties.
	The extragalactic neutrino flux from dark matter $\Phi^{\textrm{Cosm.}}$ is expected to be isotropic and to have a red-shifted decay spectrum in energy. This flux is calculated adopting the $\Lambda$CDM cosmological model with parameters from Ref.~\cite{Ade:2015xua}.
	The total signal flux is computed as the sum of both fluxes assuming that a single dark matter particle constitutes the observed dark matter in the universe. Additionally, neutrino mixing is applied with parameters from Ref.~\cite{Patrignani:2016xqp}, the effects are shown in Fig.~\ref{fig:decay_spectra_details}. The total flux depends on two fit parameters: the mass $m_{\textrm{DM}}$, which determines the energy cut-off, and the lifetime $\tau_{\textrm{DM}}$ of the dark matter particle, which determines the normalization. Explicitly, it is given by
	\begin{eqnarray}
	%\label{eq:1}
	\frac{d\Phi}{dE_{\nu}} & = & \frac{1}{4\pi m_{\textrm{DM}} \tau_{\textrm{DM}}} \left( \frac{d\Phi^{\textrm{Halo}}}{dE_{\nu}} +\frac{d\Phi^{\textrm{Cos.}}}{dE_{\nu}}  \right) ,\\ 
	\frac{d\Phi^{\textrm{Halo}}}{dE_{\nu}} & = & \frac{dN_{\nu}}{dE_{\nu}} \int \limits_0^{\infty} \rho_{\textrm{DM}}(r(s)) \, ds , \nonumber\\
	\frac{d\Phi^{\textrm{Cos.}}}{dE_{\nu}} & = & \frac{\Omega_{\textrm{DM}} \rho_{\textrm{c}}}{H_0} \int \limits_0^{\infty} \frac{dN_{\nu}}{d(E_{\nu}(1+z))}\frac{dz}{\sqrt{\Omega_{\Lambda}+\Omega_{\textrm{m}}(1+z)^3}}.\nonumber
	% \frac{d\Phi^{\textrm{Cosm.}}}{dE_{\nu}} & = & \frac{\Omega_{\textrm{DM}} \rho_{\textrm{c}}}{H_0} \int \limits_0^{\infty} \frac{1}{\sqrt{\Omega_{\Lambda}+\Omega_{\textrm{m}}(1+z)^3}} \frac{dN_{\nu}}{dE_{\nu}|_{1+z}} dz.\nonumber
	\label{eq:DM_fluxes}
	\end{eqnarray}
	
	\subsection{Likelihood analysis}
	
	In order to find the combination of the flux components that describe the data best, a forward-folding likelihood fit is performed. Flux templates, as a function of the fit parameters, are generated from a dedicated simulation of the detector response (see Refs.~\cite{Aartsen:2016xlq, Aartsen:2014muf} for more details) and then compared to the observed event distributions in reconstructed energy $E$, right-ascension $\alpha$, and zenith angle $\theta$.
	Given a set of observed events, $N$, and the predicted number of events, $\mu_i(\xi)$, the Poisson likelihood is calculated and the fit parameters $\xi$ are optimized, namely,
	\begin{equation}
	\label{eq:L_binned}
	L ( N ; \xi ) =\prod_{i=1}^{\textrm{bins}} \textrm{P}_{\textrm{Poisson}}(n_{i};\mu_{i}(E_{j},\alpha_{j},\theta_{j}; \xi)).
	\end{equation}
	While a binned likelihood method
	is used in the analysis of the track-like events, an unbinned approach is used in the analysis of the cascade sample, which corresponds to the limit of infinitesimal bin size.
	
	To quantify the statistical significance of the best fit result, a test statistic is defined as the ratio of the maximum likelihood values for the background-only case (atmospheric and astrophysical fluxes) and for the back\-ground-plus-signal case (i.e., including the additional flux from dark matter decay), namely:
	\begin{equation}
	\label{eq:TS}
	\textrm{TS}:= 2 \times \log\left(\frac{L(\hat{\phi}_{\textrm{atm.}}, \hat{\phi}_{\textrm{astro}},\hat{\gamma},\hat{m_{\textrm{DM}}},\hat{\tau_{\textrm{DM}}})}{L(\hat{\hat{\phi}}_{\textrm{atm.}},\hat{\hat{\phi}}_{\textrm{astro}},{\hat{\hat{\gamma}}},{\tau_{\textrm{DM}}}=\infty)}\right) \geq 0.
	\end{equation}
	Since the signal-plus-background case has additional degrees of freedom (four vs. two physical fit parameters), the TS value will always be positive. The observed TS value of the best-fit result is then compared to pseudo-experiments of the background and different signal hypotheses to construct confidence intervals.
	
	\subsection{Systematics}
	The systematic uncertainties of the two analyses arise from the modelling of the dark matter halo, the detector and the background fluxes.
	The dominant systematic uncertainty is the poorly understood dark matter distribution in our galactic halo. To investigate the resulting effect, the Burkert halo parameters are varied within intervals of one standard deviation while keeping their correlation fixed, by selecting $\beta_2=-0.5$ (see discussion in Ref.~\cite{Nesti:2013uwa}). In addition, the impact of a different halo profile, namely the Navarro-Frenk-White~\cite{Navarro:1995iw,Navarro:1996gj} profile, with best-fit parameters from Ref.~\cite{Nesti:2013uwa}, on the fit results is studied. The total effect of these halo model variations on the derived lifetime limit is $\pm10\%$. This value is consistent across all the masses and decay channels and between the two analyses. The uncertainty on the extragalactic flux component, which arises from the average extragalactic dark matter density, is on the order of a few percent~\cite{Ade:2015xua} and is thus not considered here. 
	
	Detector simulation and background flux uncertainties are treated differently between the two analyses. In the analysis of track-like events, several nuisance parameters are fitted simultaneously in order to absorb deviations from the baseline expectation (see Ref.~\cite{Aartsen:2016xlq} for more details). They include the normalization of the prompt atmospheric flux, cosmic-ray flux model uncertainties, relative contribution from pion and kaon decays to the atmospheric fluxes, optical properties of the glacial ice, and the optical efficiency of the detector.

	In the analysis of the cascade-like events, prompt atmospheric flux uncertainties~\cite{Aartsen:2014muf}, errors in the event reconstruction due to ice model uncertainties~\cite{Aartsen:2017eiu}, a 10\% uncertainty on the optical efficiency of the detector, and the impact of the finite simulation statistics are taken into account. The data are reanalyzed under different assumptions within the systematic uncertainties and the spread of the resulting limits is taken as the overall systematic uncertainty.
	
	\section{Results}
	\subsection{Fit results}
	
	To address the question of whether the observed flux of cosmic neutrinos can be described significantly better by including a component from decaying dark matter, the hard decay channels $DM \rightarrow H + \nu $ (cascades) and $DM \rightarrow Z + \nu $ (tracks) are fitted to the respective data. A dark matter signal would be expected to show up in both analyses.
	Also note, that the observable energy distributions are smeared out due to the limited detector resolution and the cosmological red-shift. It is therefore sufficient to fit these single decay channels in order to test whether a contribution from dark matter is present and multiple tests are not necessary.
	The obtained best-fit results and the corresponding p-values with respect to the background only hypothesis are listed in Table \ref{tab:best_fits}. The fits of the background-only hypothesis agree well with the results in Refs.~\cite{Aartsen:2016xlq} and~\cite{Aartsen:2014muf}. Small differences arise due to a different choice of bins (tracks) and the altered selection (cascades).
	
\begin{figure}[t]%[htbp]
		\includegraphics[width=.48\textwidth,left]{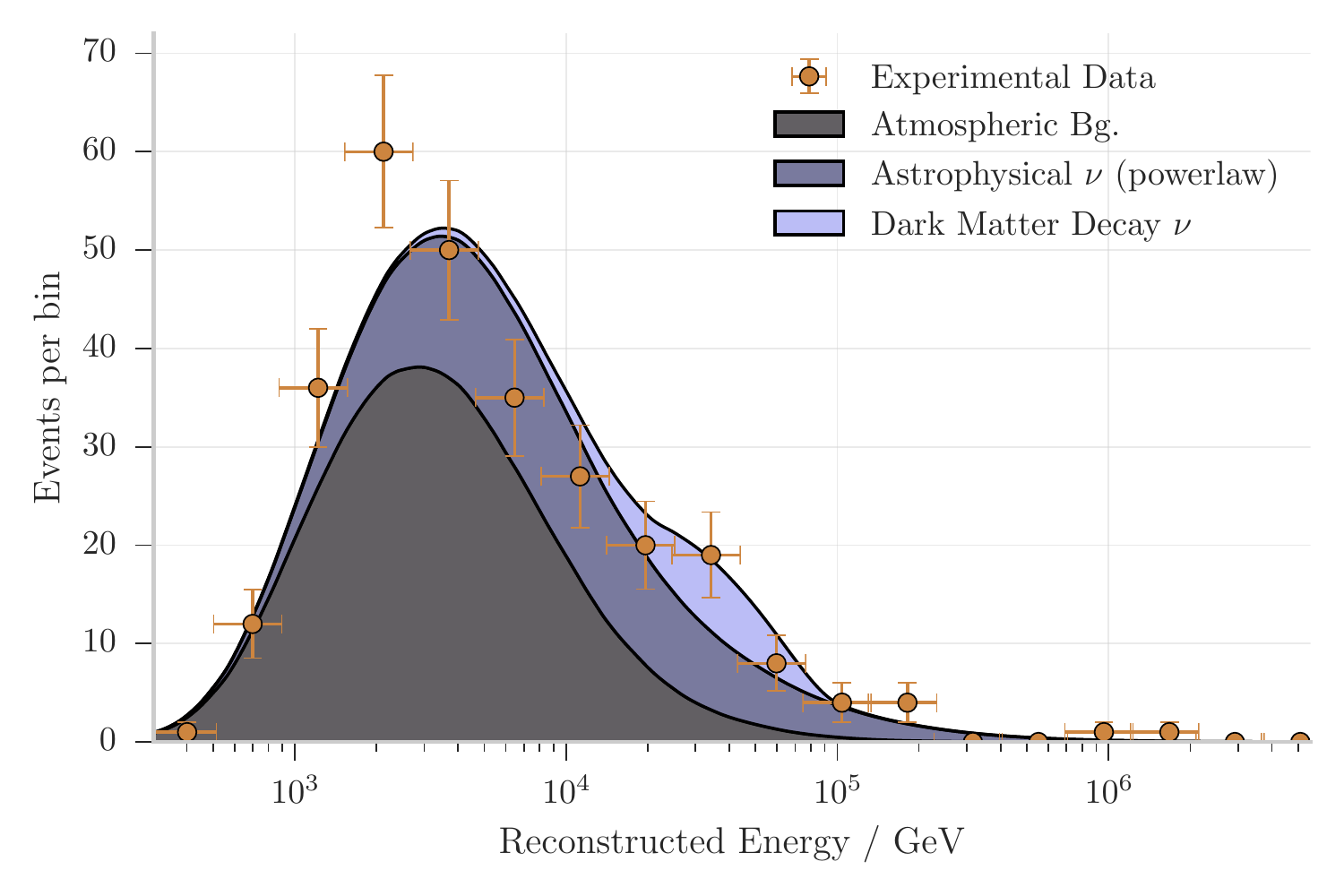}  
		\protect\caption{Cascade Analysis: Best-fit energy distribution for the signal hypothesis (components stacked to illustrate the dark matter component), with the best fit parameters listed in Table \ref{tab:best_fits}. The fit is performed on un-binned data, but for visualization purposes a binning is applied in the figure.}
		\label{fig:bf_energydistribution_cascades}
	\end{figure}

\begin{figure}[t]%[htbp]
		\includegraphics[width=.48\textwidth,left]{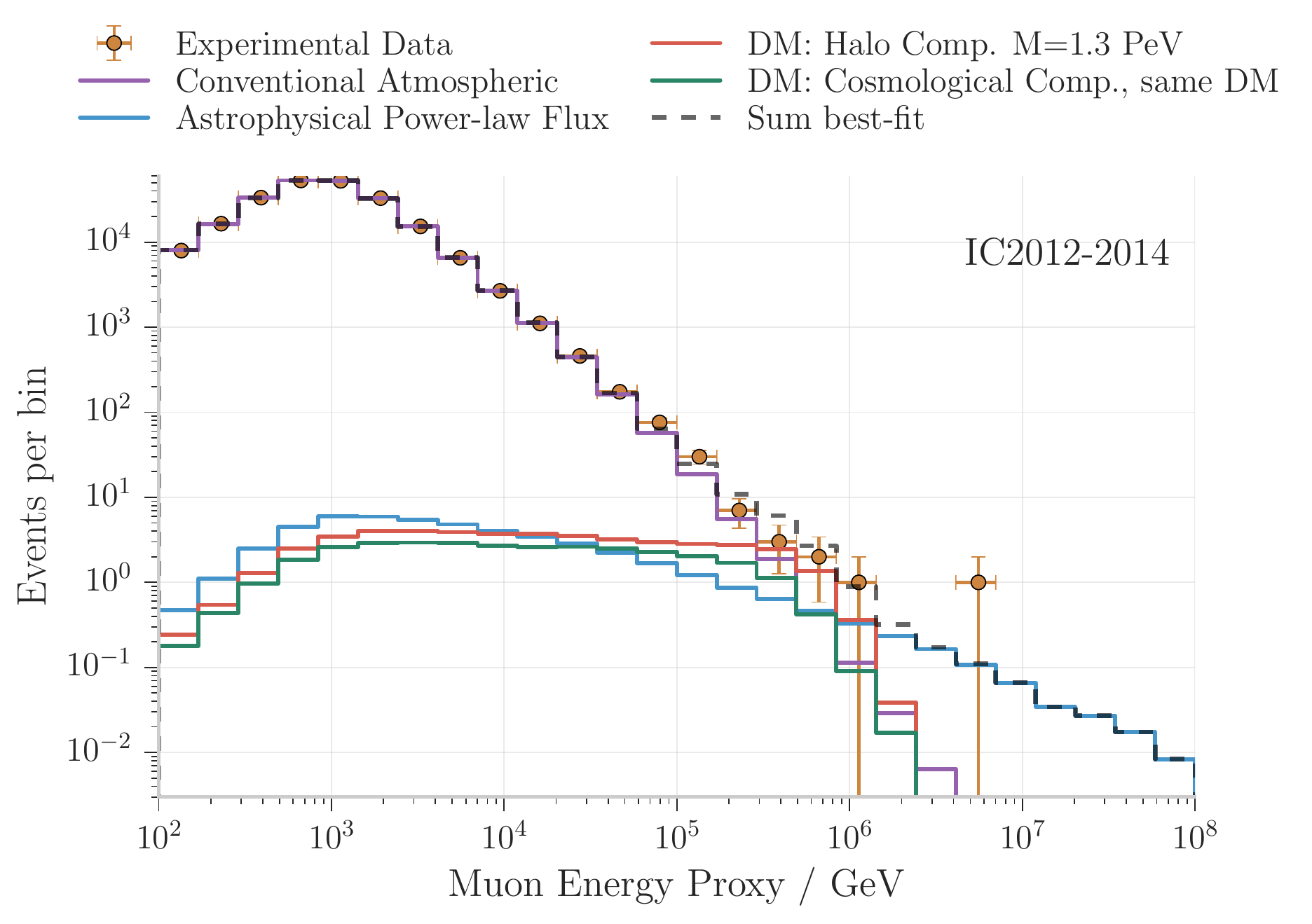}  
		\protect\caption{Track Analysis: Best-fit energy distribution. While the low-energy events are well described by the conventional atmospheric component, the high-energy events are modelled by a combination of a weak diffuse astrophysical flux and a component from decaying dark matter (best-fit parameters in Table \ref{tab:best_fits}). 
			The figure shows data recorded between 2012 and 2014 as they are based on the same energy estimator (see~\cite{ABBASI2013190} for more details). The remaining years are fitted simultaneously but are not shown here.}
		\label{fig:bf_energydistribution_tracks}
	\end{figure}

	\begin{table}[th]%[bhtp]
		\centering
		\caption{\label{tab:best_fits} Best-fit results assuming the decay channels $DM \rightarrow H + \nu $ (cascades) and $DM \rightarrow Z + \nu $ (tracks). Background p-values are stated in brackets.}
		\begin{tabular*}{\columnwidth}{@{\extracolsep{\fill}}lcccc}
			%\begin{tabular}{lcccc}
			\hline
			& \multicolumn{2}{c}{Tracks} & \multicolumn{2}{c}{Cascades}\\
			& Bg. 		& Signal+Bg. 	& Bg. 	& 	Signal+Bg. \\ 
			\hline
			$m_{\textrm{DM}}\,/\,\text{PeV}$ & - & 1.3 & - & 0.1\\
			$\tau_{\textrm{DM}}\,/\,10^{27}\text{s}$ & - & 22 & - & 8.3\\
			Astroph. norm.\footnotemark & 0.97 & 0.16 & 2.15 & 1.62\\
			Spectr. index  & 2.16 & 1.99 & 2.75 & 2.81\\
			$\textrm{TS} = 2\times \Delta \textrm{LLH}$ & \multicolumn{2}{c}{$6.7\,(p=0.035$)} &  \multicolumn{2}{c}{ $3.4\,(p=0.55)$}\\
			\hline
		\end{tabular*}
		\footnotetext{Normalization in units of $10^{-18}$GeV$^{-1}$cm$^{-2}$sr$^{-1}$s$^{-1}$.}
	\end{table}
	The corresponding best-fit distributions in reconstructed energy are shown in Figures~\ref{fig:bf_energydistribution_cascades} and \ref{fig:bf_energydistribution_tracks} together with the experimental data. Note that different energy estimators are used in the sub-samples (data-taking seasons) of the track analysis~\cite{ABBASI2013190}. It is therefore not possible to show the experimental data in one histogram. 
	
	\subsection{Interpretation of the fit results}
\begin{figure}[ht]
		\begin{centering}
			\includegraphics[width=.48\textwidth]{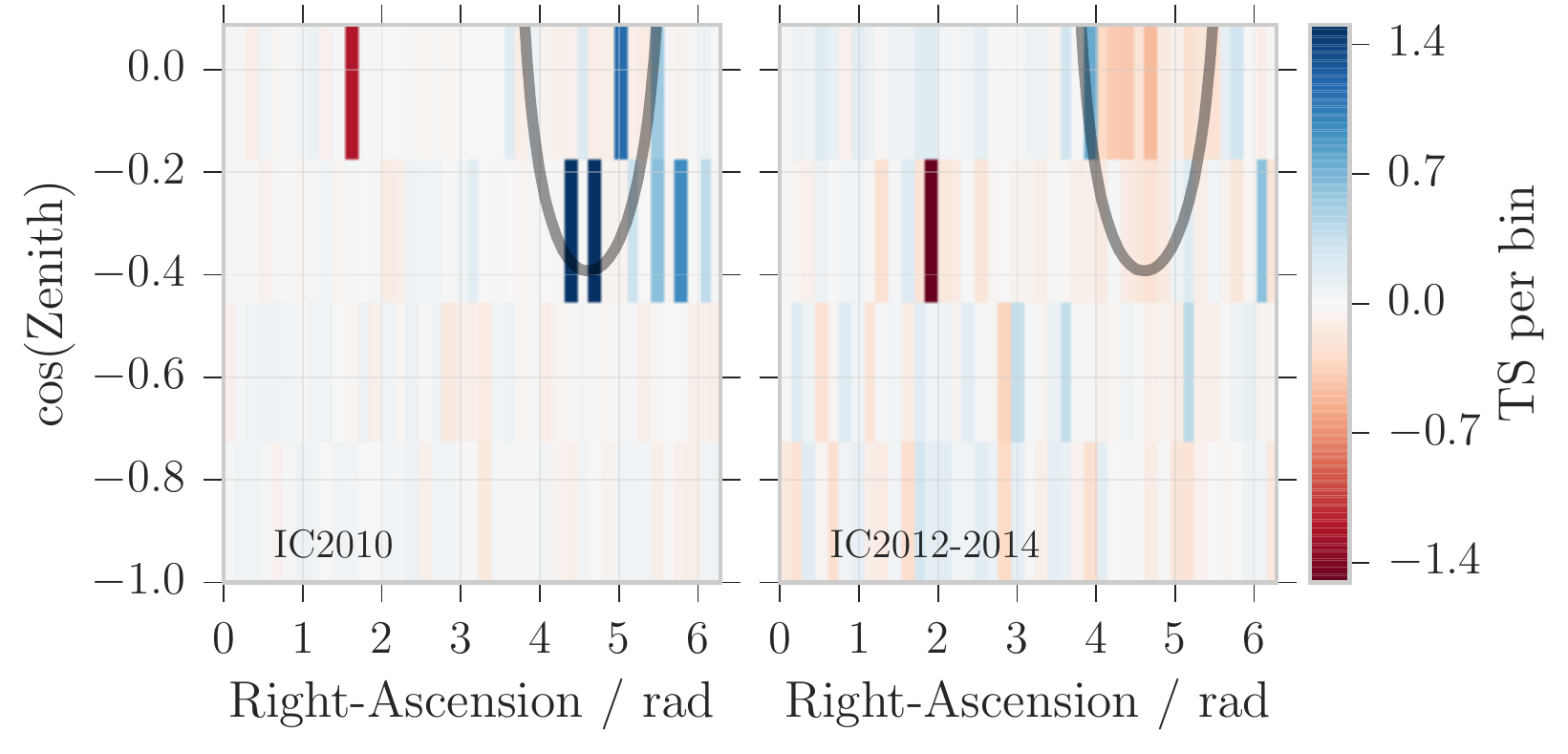}
			\par\end{centering}
		\protect\caption{Track Analysis: TS per bin to illustrate the time-dependency of the fit result: blue bins show agreement with the signal hypothesis, red bins favour a purely diffuse astrophysical flux. The gray line indicates the direction where most of the dark matter signal is expected (line-of-sight integral at half of the central value).}
		\label{fig:astro_LLHperBin}
	\end{figure}

	Although the best-fit result in both analyses includes a non-zero dark matter component, the results are not significant (as both p-values are above $1\%$). More degrees of freedom in the modelling of the astrophysical flux, e.g. adding a second component, would further reduce the significance. Thus, the result is not interpreted as a signal of dark matter decay.
	% Also, no agreement is found between the dark matter fit parameters in the two performed analyses; The best-fit values in Table~\ref{tab:best_fits} are disfavoured by the other analysis with more than one standard deviation, respectively.
	Furthermore, a dark matter signal should be constant in time but the fit of the track-like events shows fluctuations; see Fig.~\ref{fig:astro_LLHperBin}: While those bins contributing most strongly in the fit to the data from the first three years (e.g., 2010) coincide with the approximate direction of the dark matter halo, such a correlation is disfavoured by the data from 2012-2014. 
	
\begin{figure}[ht]%htbp]
		\begin{centering}
			\includegraphics[width=.45\textwidth]{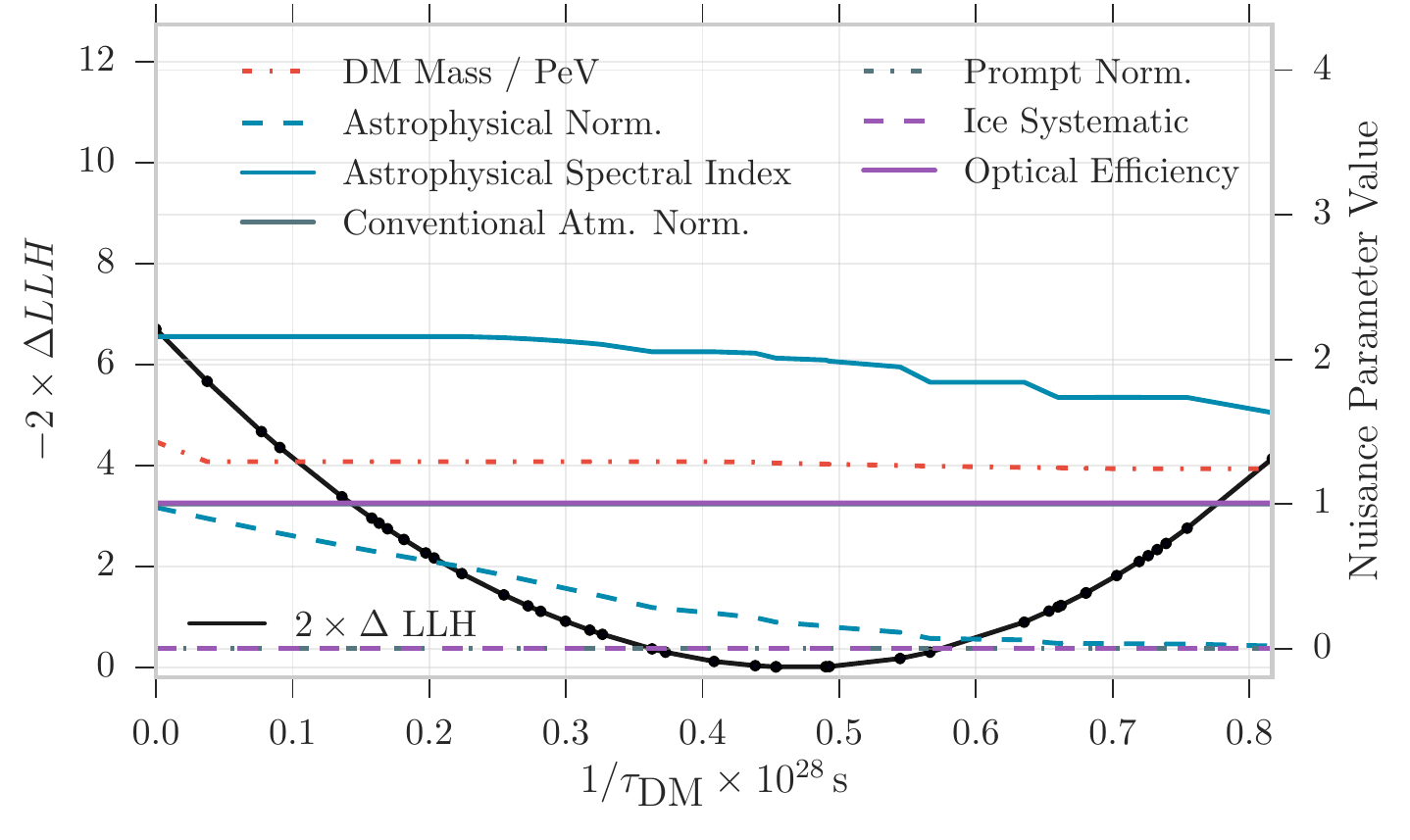}
			\includegraphics[width=.45\textwidth]{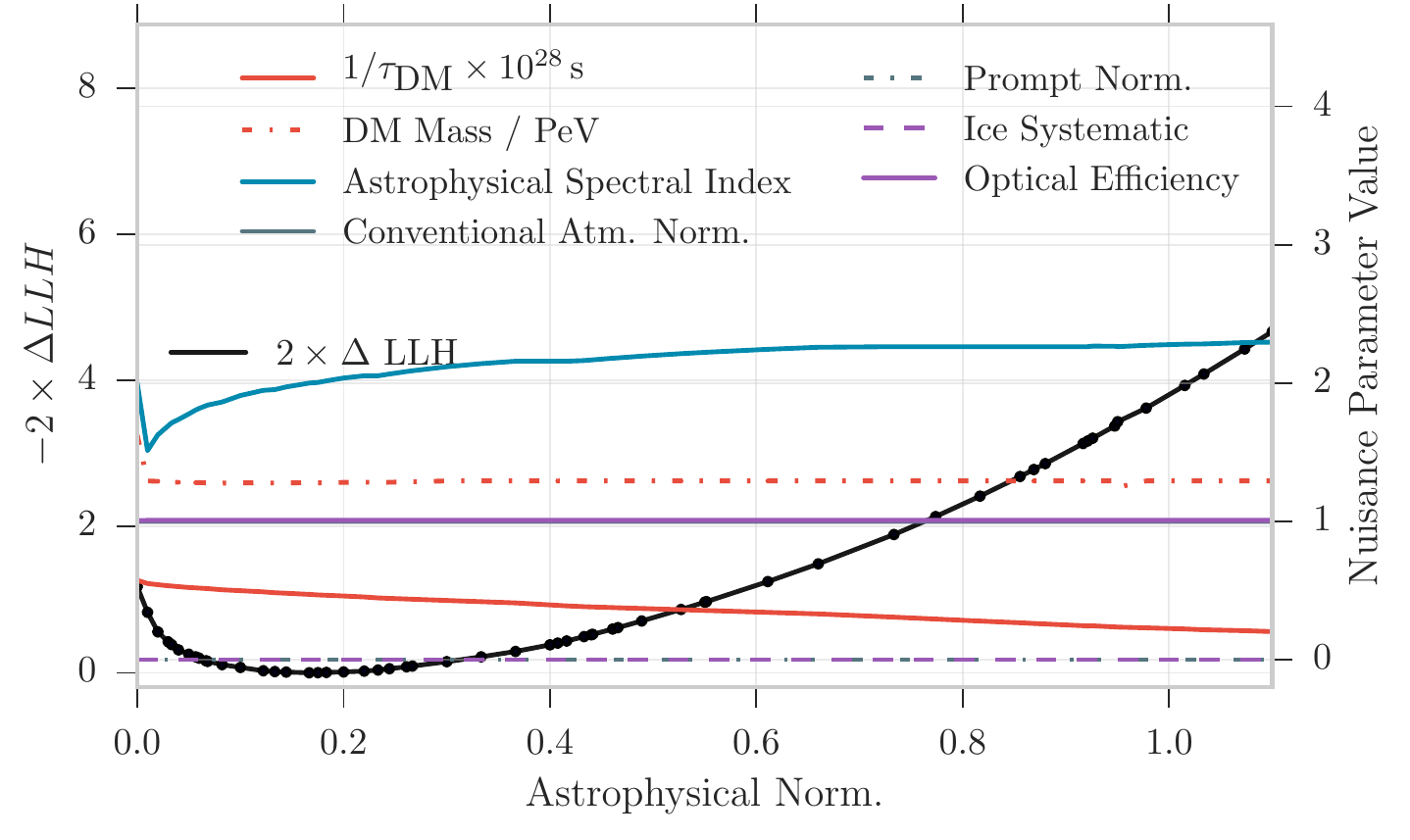}
			\par\end{centering}
		\protect\caption{Track analysis: Profile likelihood scans of the inverse dark matter lifetime (proportional to the signal strength) and the diffuse astrophysical flux normalization in units of $10^{-18}$GeV$^{-1}$cm$^{-2}$sr$^{-1}$s$^{-1}$.}
		\label{fig:astro_interplay}
	\end{figure}

	Another interesting observation is the interplay of the diffuse astrophysical flux and the dark matter component in the fit of track-like events: Fig.~\ref{fig:astro_interplay} shows the profile likelihood of the respective normalizations together with the fit result of other selected parameters. The best-fit astrophysical normalization is significantly reduced compared to previous results~\cite{Aartsen:2016xlq}. A dark matter only scenario, where the normalization of the astrophysical flux is zero, is however disfavoured by $2\Delta LLH\simeq1$ compared to the best-fit point.
	As expected, the best-fit dark matter mass that induces a cut-off in the energy spectrum is found to be independent of the diffuse astrophysical normalization while the dark matter normalization is anti-correlated.
	
	\subsection{Lifetime limits}
\begin{figure}[ht]%[ht]
		\includegraphics[width=.47\textwidth,left]{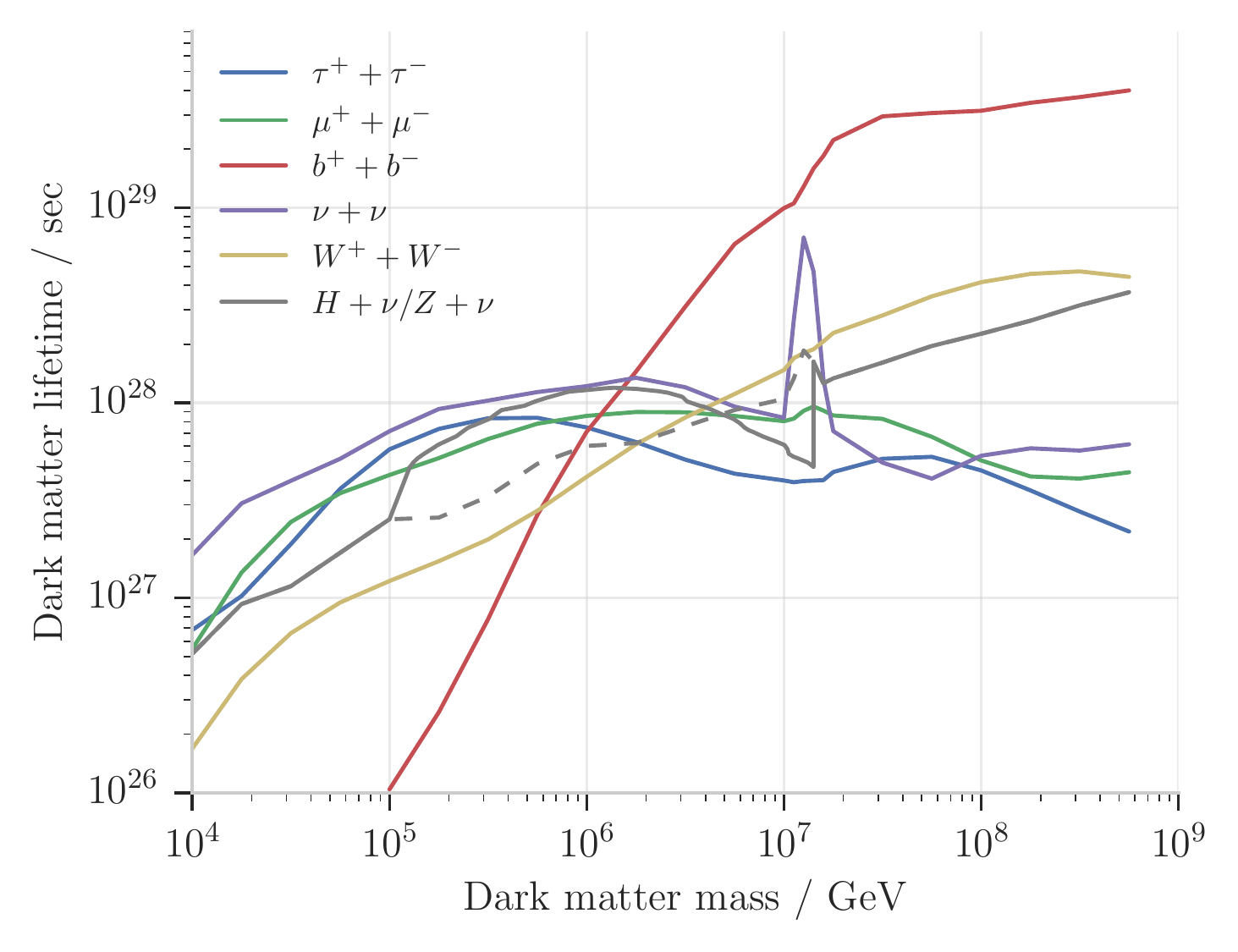}
		\protect\caption{Dark matter lifetime limits for all considered decay channels. For the $Z + \nu / H + \nu$ channel, the limit was combined (solid grey line) as described in the text. Between $m_{\textrm{DM}}\sim10^5\,\si{\GeV}$ and $m_{\textrm{DM}}=1.5\times10^7\,\si{\GeV}$ the limit is obtained from the more sensitive track analysis. The limit from the cascade analysis is shown as a dashed line and turns out to be stronger above $m_{\textrm{DM}}\sim \SI{5e7}{\GeV}$.}
		\label{fig:limits}
	\end{figure}

	Since no significant dark matter signal is observed, lower limits on the lifetime of the dark matter particle (corresponding to upper limits on its signal strength) are derived. In order to combine the two analyses, the lower limit on the lifetime is extracted from the respective analysis with the better sensitivity (median limit obtained from background pseudo-experiments) at each dark matter mass. 
	The hard decay channels ${\rm Z}+\nu$ (track analysis) and ${\rm H}+\nu$ (cascade analysis) are treated as the same channel because the resulting neutrino spectra are indistinguishable within energy resolutions. Further, limits for the decay channels $\nu\bar{\nu}$, $\tau^{+}\tau^{-}$, $\mu^{+}\mu^{-}$, $W^{+}W^{-}$ and $b\bar{b}$ are calculated only in the cascade analysis because the energy resolution of the track analysis is not sufficient to differentiate those channels from each other. The resulting lower limits on the dark matter lifetime are shown in Fig.~\ref{fig:limits}. 
	Note that for the $b\bar{b}$ decay channel, the lower limit on the lifetime increases steeply with the dark matter mass because QCD fragmentation generates a soft tail of low-energy neutrinos (see Fig.~\ref{fig:decay_spectra}) which become increasingly relevant for large dark matter mass. Furthermore, no limit on the lifetime is calculated in this channel for $m_{\textrm{DM}}$ below $10^5\,\,\si{\GeV}$ because the resulting decay spectrum becomes similar to the atmospheric background fluxes and the respective uncertainties would have a major effect on the obtained limit. The enhanced limits at $m_{\textrm{DM}}\sim 10^7~\si{\GeV}$, correspond to the non-observation of electron neutrino events from the expected Glashow resonance~\cite{PhysRev.118.316}. 
	For the track-like sample, all nuisance parameters are fitted to their expectation values within one standard deviation, and the effect on the signal hypothesis is found to be negligible. For the cascade-like sample, the overall impact of the systematics is roughly 10-15\% for dark matter masses below 5 PeV and 1\% for those above it. The limits shown here include a degradation due to $1\sigma$ systematic variation. 
	
	\begin{figure}[ht]%[ht]
		\includegraphics[width=.47\textwidth,left]{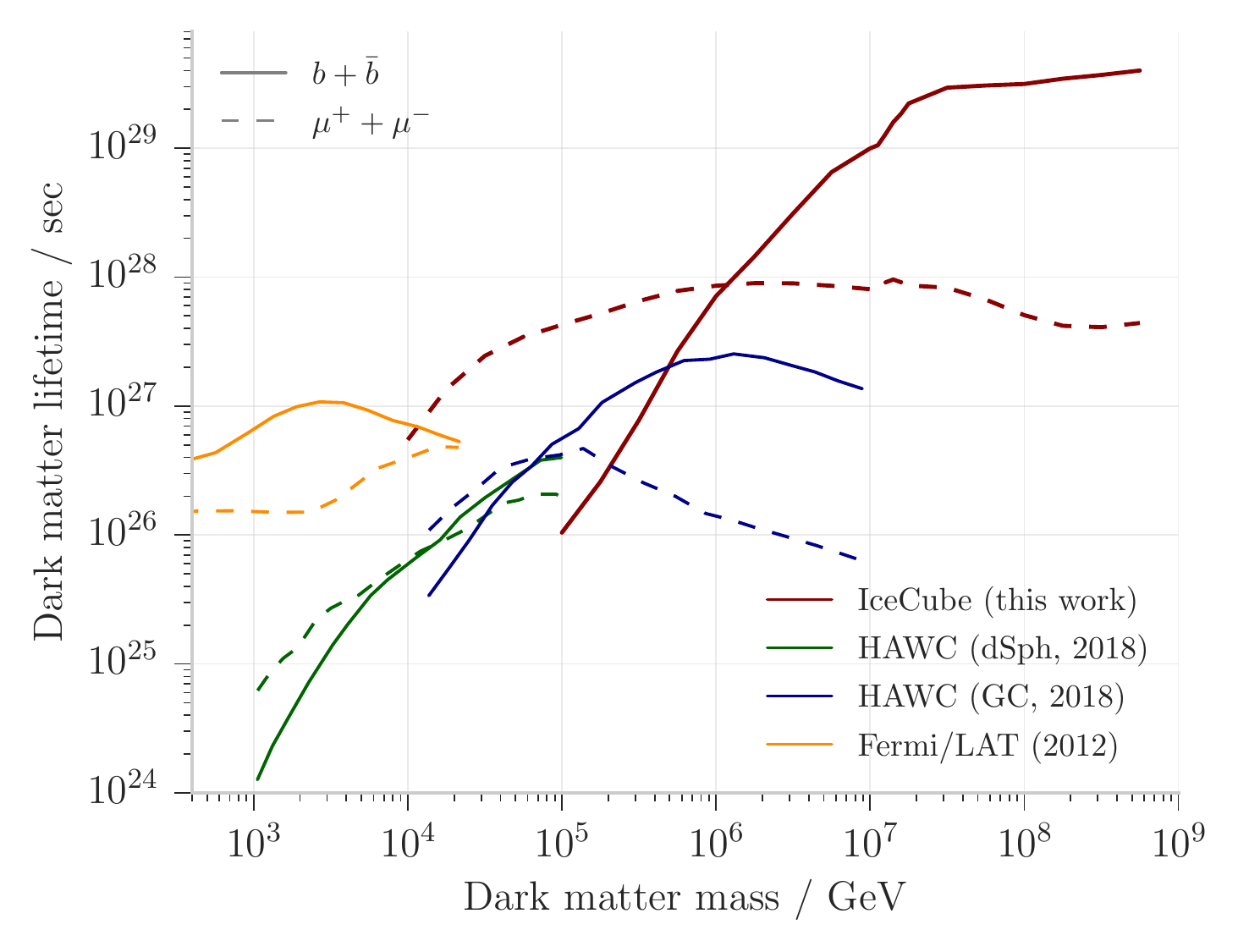}
		\protect\caption{Comparison of the lower lifetime limits with results obtained from gamma-ray telescopes: HAWC (Dwarf Spheroidal Galaxies)~\cite{HAWC_dm}, HAWC (Galactic Halo/Center)~\cite{HAWC_dm_GC_2018} and Fermi/LAT~\cite{Ackermann:2012rg}.}
		\label{fig:limit_comp}
	\end{figure}
	
	\section{Conclusions}
	
	Two analyses on statistically independent datasets searching for a contribution from decaying dark matter to the astrophysical neutrino flux have been presented. It has been shown that the observed high-energy neutrino flux can be described equally well by a combination of a dark matter component and a diffuse astrophysical flux with a power-law energy spectrum. However, neither analysis identified a significant dark matter excess in the data, and models in which the cosmic neutrinos flux arises entirely from dark matter decay are disfavoured. 
	
	From the non-observation of a dark matter signal, lower limits are set on the lifetime of dark matter particles with mass above $10^{4}\,\si{\GeV}$. For such heavy particles these limits are presently the strongest on the dark matter lifetime (see Fig. \ref{fig:limit_comp}).
	
	\begin{acknowledgements}
		The IceCube Collaboration designed, constructed and now operates the IceCube Neutrino Observatory. Data processing and calibration, Monte Carlo simulations of the detector and of theoretical models, and data analyses were performed by a large number of collaboration members, who also discussed and approved the scientific results presented here. The main authors of this manuscript were Hrvoje Dujmovic and J\"oran Stettner. It was reviewed by the entire collaboration before publication, and all authors approved the final version of the manuscript.
		
		We acknowledge support from the following agencies:
		
		USA -- U.S. National Science Foundation-Office of Polar Programs,
		U.S. National Science Foundation-Physics Division,
		Wisconsin Alumni Research Foundation,
		Center for High Throughput Computing (CHTC) at the University of Wisconsin-Madison,
		Open Science Grid (OSG),
		Extreme Science and Engineering Discovery Environment (XSEDE),
		U.S. Department of Energy-National Energy Research Scientific Computing Center,
		Particle astrophysics research computing center at the University of Maryland,
		Institute for Cyber-Enabled Research at Michigan State University,
		and Astroparticle physics computational facility at Marquette University;
		Belgium -- Funds for Scientific Research (FRS-FNRS and FWO),
		FWO Odysseus and Big Science programmes,
		and Belgian Federal Science Policy Office (Belspo);
		Germany -- Bundesministerium f\"ur Bildung und Forschung (BMBF),
		Deutsche Forschungsgemeinschaft (DFG),
		Helmholtz Alliance for Astroparticle Physics (HAP),
		Initiative and Networking Fund of the Helmholtz Association,
		Deutsches Elektronen Synchrotron (DESY),
		and High Performance Computing cluster of the RWTH Aachen;
		Sweden -- Swedish Research Council,
		Swedish Polar Research Secretariat,
		Swedish National Infrastructure for Computing (SNIC),
		and Knut and Alice Wallenberg Foundation;
		Australia -- Australian Research Council;
		Canada -- Natural Sciences and Engineering Research Council of Canada,
		Calcul Qu\'ebec, Compute Ontario, Canada Foundation for Innovation, WestGrid, and Compute Canada;
		Denmark -- Villum Fonden, Danish National Research Foundation (DNRF);
		New Zealand -- Marsden Fund;
		Japan -- Japan Society for Promotion of Science (JSPS)
		and Institute for Global Prominent Research (IGPR) of Chiba University;
		Korea -- National Research Foundation of Korea (NRF);
		Switzerland -- Swiss National Science Foundation (SNSF).
	\end{acknowledgements}
	
	%The Revtex-style  (PRL) was used for citations. But I went through all journals and replaced with the abbreviations as requested by EPJ guidelines.
	
	%Finally, the entries from the (once compiled) .bbl file were copied here.
	%\bibliographystyle{revetex_41_citestyle}
	%\bibliography{bibliography}
	
	% Extra hack: Remove the NoStop entries from the bbl file to remove the trailing . after reference entries by hand :-D 
	%\include{compiled_bbl}
	%merlin.mbs apsrev4-1.bst 2010-07-25 4.21a (PWD, AO, DPC) hacked
	%Control: key (0)
	%Control: author (8) initials jnrlst
	%Control: editor formatted (1) identically to author
	%Control: production of article title (-1) disabled
	%Control: page (0) single
	%Control: year (1) truncated
	%Control: production of eprint (0) enabled
	%
	
\end{document}